\documentclass[amsmath,amssymb,showpacs,twocolumn,superscriptaddress,pr]{revtex4-1}
\usepackage{graphicx,bm,color,subfigure}
\usepackage{appendix}

\begin{document}

\title{Intrinsic topological superconductivity with exactly flat surface bands in the quasi-one-dimensional A$_2$Cr$_3$As$_3$ (A=Na, K, Rb, Cs) superconductors }

\author{Cheng-Cheng Liu}
\affiliation{School of Physics, Beijing Institute of Technology, Beijing 100081, China}

\author{Chen Lu}
\affiliation{School of Physics and Technology, Wuhan University, Wuhan 430072, China}

\author{Li-Da Zhang}
\affiliation{School of Physics, Beijing Institute of Technology, Beijing 100081, China}

\author{Xianxin Wu}
\affiliation{Institut f\"ur Theoretische Physik und Astrophysik, Julius-Maximilians-Universit\"at W\"urzburg, 97074 W\"urzburg,
  Germany}

\author{Chen Fang}
\affiliation{ Institute of Physics, Chinese Academy of Sciences, Beijing 100190, China}

\author{Fan Yang}
\email{yangfan\_blg@bit.edu.cn}
\affiliation{School of Physics, Beijing Institute of Technology, Beijing 100081, China}

\begin{abstract}
A spin-U(1)-symmetry protected momentum-dependent integer-$Z$-valued topological invariant is proposed to time-reversal-invariant (TRI) superconductivity (SC) whose nonzero value will lead to exactly flat surface band(s). The theory is applied to the weakly spin-orbit coupled quasi-1D A$_2$Cr$_3$As$_3$ (A=Na, K, Rb, Cs) superconductors family with highest $T_c$ up to 8.6 K with $p_z$-wave pairing in the $S_z=0$ channel. It's found that up to the leading atomic spin-orbit-coupling (SOC), the whole (001) surface Brillouin zone is covered with exactly-flat surface bands, with some regime hosting three flat bands and the remaining part hosting two. Such exactly-flat surface bands will lead to very sharp zero-bias conductance peak in the scanning tunneling microscopic spectrum. When a tiny subleading spin-flipping SOC is considered, the surface bands will only be slightly split. The application of this theory can be generalized to other unconventional superconductors with weak SOC, particularly to those with mirror-reflection symmetry.
\end{abstract}


\maketitle

\textit{\textcolor{black}{Introduction.---}} Topological superconductivity (TSC) has aroused great interests in the past decades\cite{TQC1,TQC2}. The key feature of TSC lies in the presence of gapless Majorana Fermions at the end (for 1D), edge (for 2D) or surface (for 3D)\cite{Read,Kitaev1,Kitaev2,Schnyder,Fu2008,Qi,Ryu,Sasaki2011,Yang2014,Chiu2016,Chiu2014,Micklitz2017,Kobayashi2018,Sumita2018,Tanaka_RPP}. In 1D, it's proposed that an effective $p$-wave TSC realized via Rashba spin-orbit-coupling (SOC) with Zeeman coupling\cite{Dassarma} can accommodate Majorana end  state, detected by the scanning tunneling microscope (STM) as a pronounced zero-bias conductance peak (ZBCP)\cite{Yazidani}. However, in 2D or 3D, the dispersion of the Majorana bands on the edge or surface will broaden the bands and lead to weak\cite{Yamashiro,Tanaka1,Tanaka2} or no ZBCP\cite{Yamakage,Asano} in the STM. Therefore, the experimental identification of TSC in higher than 1D is still a challenge.

Here we investigate the evolution of the isolated end states of a 1D $p$-wave superconductor when a weak imposed three dimensionality expands them into several branches of surface bands with dispersion. It's interesting to ask: is it possible to realize a quasi-1D TSC protected by some symmetries which hosts dispersionless surface band(s)? Here we propose a new class of TSC protected by the time-reversal (TR) and the spin-U(1) symmetry (SUS), which hosts exactly flat surface band(s). Different from conventional nodal-line TSC\cite{Ryu,Tanaka2010,Schnyder2011,Brydon2011,Schnyder2012, Tanaka1995}, the flat surface band(s) here doesn't rely on the presence of the nodal line, and the whole surface Brillouin zone (BZ) can be covered by flat band(s), causing sharp ZBCP in the STM. Furthermore, we propose that the recently synthesized quasi-1D  A$_2$Cr$_3$As$_3$ (A=Na, K, Rb, Cs) family \cite{Bao,Tang1,Tang2,Mu} with predicted $p_z$-wave pairing symmetry\cite{Wu,Zhang} belong to this TSC class, up to the leading SOC.

The low-energy degrees of freedom in the A$_2$Cr$_3$As$_3$ family are dominated by the Cr-3d orbitals\cite{Cao1,Hu1}, which are expected to be strongly-correlated\cite{Wu,Zhang,Zhou1,Zhou2,Dai}, supported by various experiments\cite{Imai,Zheng1,Taddei,Raman,ARPES}, implying an electron--interaction-driven pairing mechanism. Diverse experiments\cite{Bao, Tang1,Tang2,Imai,Zheng1,uncsc1,uncsc2,uncsc3,uncsc4, Zheng2019} have revealed unconventional pairing states, particularly with line nodes\cite{uncsc1,Tang1} and possibly triplet pairings\cite{Bao, Tang1,Tang2} in the system. Symmetry analysis suggests that the leading SOC in the A$_2$Cr$_3$As$_3$ family is the atomic SOC conserving the SUS\cite{Wu}, and combined weak- and strong- coupling approaches have predicted TRI $p_z$-wave triplet pairing with line nodes with $S_z=0$ component\cite{Wu,Zhang}, belonging to the symmetry class required here up to the leading SOC.

In this Letter, we provide topological invariant associated with flat surface band(s) under combined TR+SUS symmetries for TSCs, and apply it to the quasi-1D A$_2$Cr$_3$As$_3$ family. As a result, the momentum-dependent topological invariant $Z(k_x,k_y)$ for A$_2$Cr$_3$As$_3$ is nonzero all over the $(k_x,k_y)$-plane, with different regimes covered with different $Z$. Consequently, different regimes on the surface BZ are covered with different nonzero numbers of flat bands, causing sharp ZBCP in the STM. Our proposal provides smoking-gun evidence for experimental identification of such quasi-1D $p_z$-wave superconductors as the A$_2$Cr$_3$As$_3$ with weak SOC conserving the SUS.

\textit{\textcolor{black}{SUS protected Topological invariant.---}} Let's consider a multi-band TRI superconductor with SUS, whose Bogolubov-de Gennes (BdG) Hamiltonian is \begin{equation}\label{origin_H}
H=\sum_{\mathbf{k\alpha\sigma}}\varepsilon_{\mathbf{k}\alpha\sigma}c^{\dagger}_{\mathbf{k}\alpha\sigma}c_{\mathbf{k}\alpha\sigma}+
\sum_{\mathbf{k}\alpha\beta}\left[\Delta_{\alpha\beta}(\mathbf{k})c^{\dagger}_{\mathbf{k}\alpha\uparrow}c^{\dagger}_{\mathbf{-k}\beta\downarrow}+h.c.\right].
\end{equation}
Here $\alpha/\beta=1,\cdots,N_{\alpha}$ represent the band indices, and $\sigma$ labels spin. The TR symmetry requires $\varepsilon_{\mathbf{k}\alpha\uparrow}=\varepsilon_{-\mathbf{k}\alpha\downarrow}$ and $\Delta_{\mathbf{k}}=\Delta_{\mathbf{k}}^{\dagger}$. This BdG Hamiltonian can be
written in the particle-hole symmetric (PHS) 4-component Nambu representation as $H=\frac{1}{2}\sum_{\mathbf{k}}\Psi_{\mathbf{k}}^{\dagger}\mathcal{H}\Psi_{\mathbf{k}}$, with $\Psi^{\dagger}_{\mathbf{k}}=\left(\mathbf{c_{\mathbf{k}\uparrow}^{\dagger}},\mathbf{c_{\mathbf{k}\downarrow}^{\dagger}},\mathbf{c_{-\mathbf{k}\uparrow}},\mathbf{c_{-\mathbf{k}\downarrow}}\right)$, $\mathbf{c_{\mathbf{k}\uparrow}^{\dagger}}\equiv(\cdots\mathbf{c_{\mathbf{k}\alpha\uparrow}^{\dagger}}\cdots)$, and
\begin{equation}
\label{BdG}
\mathcal{H}=\left(\begin{array}{cccc}
\varepsilon_{\mathbf{k}\uparrow} & 0 & 0 & \Delta_{\mathbf{k}}\\
0 & \varepsilon_{\mathbf{k}\downarrow} & -\Delta_{-\mathbf{k}}^{T} & 0\\
0 & -\Delta_{-\mathbf{k}}^{T} & -\varepsilon_{\mathbf{k}\downarrow} & 0\\
\Delta_{\mathbf{k}} & 0 & 0 & -\varepsilon_{\mathbf{k}\uparrow}
\end{array}\right).
\end{equation}
Here $\varepsilon_{\mathbf{k}\sigma}\equiv \text{diag}(\cdots, \varepsilon_{\mathbf{k}\alpha\sigma}, \cdots)$. Note that the SUS requires the hopping blocks of the BdG matrix Eq.(\ref{BdG}) to be diagonal and the pairing blocks to be block off-diagonal. The combined TRS and PHS lead to chiral symmetry, which enables us to do the unitary transformation $\mathcal{H}\rightarrow\widetilde{\mathcal{H}}=U^{\dagger}\mathcal{H}U$ to obtain an off-diagonal Hermitian matrix $\widetilde{\mathcal{H}}$. Here the unitary matrix $U$ and the upper-right off-diagonal block $\widetilde{\mathcal{H}}_{12}$ of $\widetilde{\mathcal{H}}$ read
\begin{equation}\label{unitary}
U=\frac{1}{\sqrt{2}}\left(\begin{array}{cc}
I & I\\
\sigma_{y} & -\sigma_{y}
\end{array}\right),\widetilde{\mathcal{H}}_{12}=\left(\begin{array}{cc}
\varepsilon_{\mathbf{k}\uparrow}-i\Delta_{\mathbf{k}} & 0\\
0 & \varepsilon_{\mathbf{k}\downarrow}-i\Delta_{-\mathbf{k}}^{T}
\end{array}\right).
\end{equation}
Note that $\widetilde{\mathcal{H}}_{12}$ is block-diagonal, caused by the SUS.

Following the standard procedure introduced in Ref~\cite{Schnyder,Chiu2016,Schnyder2011}, the so-called Q-matrix is obtained
\begin{equation}\label{Q_matrix}
Q\left(\mathbf{k}\right)=I-2P\left(\mathbf{k}\right)=\left(\begin{array}{cccc}
 &  & \mathbf{q}_{1} & 0\\
 &  & 0 & \mathbf{q}_{2}\\
\mathbf{q}_{1}^{\dagger} & 0\\
0 & \mathbf{q}_{2}^{\dagger}
\end{array}\right).
\end{equation}
Here $P\left(\mathbf{k}\right)$ is the projection operator, and the $N_{\alpha}\times N_{\alpha}$ matrix $q_{1}$ ($q_{2}$) is related to the upper-left (lower-right) block of $\widetilde{\mathcal{H}}_{12}$ in Eq.(\ref{unitary}), and hence the 1-4 (2-3)-block of $\mathcal{H}$ in Eq.(\ref{BdG}). Each block of the Hamiltonian only has the chiral symmetry and belongs to the AIII class, characterized by a $Z$ topological invariant in 1D. The detailed formulae of $\mathbf{q}_{1/2}$ is provided in the Appendix \ref{appendixA}. Note that in contrast with the cases in ordinary TRI superconductor\cite{Schnyder,Chiu2016,Schnyder2011}, the extra SUS here makes the off-diagonal block $q\equiv\text{diag}(\mathbf{q}_1,\mathbf{q}_2)$ of the $Q$ matrix block-diagonalized into $\mathbf{q}_1$ and $\mathbf{q}_2$ sub-blocks.  This enables us to define the following two 1D winding number $Z_{1/2}$ for the two sub-blocks $\mathbf{q}_{1/2}$, instead of the one $Z$ defined for the whole $\mathbf{q}$ for ordinary TRI superconductors\cite{Schnyder,Chiu2016,Schnyder2011}.
\begin{equation}
\label{windingnumber}
Z_{1/2}(k_x,k_y)=\frac{i}{2\pi}\int^{\pi}_{-\pi} tr\left(\mathbf{q}_{1/2}^{\dagger}\partial_{k_z}\mathbf{q}_{1/2}\right)d k_z.
\end{equation}
Here in defining the path-dependent 1D winding number $Z_{1/2}$, we have chosen the path ``$L$''\cite{Schnyder2011} to be a vertical line passing $(k_x,k_y,0)$. Note that due to the double counting brought about by the gauge redundancy in this representation, the physical topological invariant here should be $Z_1(k_x,k_y)$, which leads to $|Z_1(k_x,k_y)|$ flat surface bands as shown in the Appendix \ref{appendixA}. The relationship between the winding numbers $Z$ and $Z_{1/2}$  is similar to that between the Chern number and the spin Chern number\cite{spin_chern_1, spin_chern_2}.

It's interesting to investigate the case of intraband-pairing limit where $\Delta_{\mathbf{k}}^{\alpha\beta}$$=$$\Delta_{\mathbf{k}}^{\alpha}\delta_{\alpha\beta}$. In this case, it's proved in Appendix \ref{appendixA} that $q_1(\mathbf{k})=\text{diag}(\cdots, e^{-i\theta_{\mathbf{k}\alpha\uparrow}},\cdots)$ with $e^{-i\theta_{\mathbf{k}\alpha\uparrow}}=\left(\varepsilon_{\mathbf{k}\alpha\uparrow}-i\Delta_{\mathbf{k}}^{\alpha}\right)/\mid\varepsilon_{\mathbf{k}\alpha\uparrow}-i\Delta_{\mathbf{k}}^{\alpha}\mid$.
Then from Eq.(\ref{windingnumber}), the winding number is obtained as
\begin{equation}\label{intraband_limit}
Z_{1}(k_x,k_y)=\frac{1}{2\pi}\sum_{\alpha}\int^{\pi}_{-\pi}d\theta_{\mathbf{k}\alpha\uparrow}=\sum_{\alpha}I_{\alpha}(k_x,k_y).
\end{equation}
Eq.(\ref{intraband_limit}) suggests that in this limit, $Z_{1}(k_x,k_y)$ is a summation of the contributions from each band $\alpha$, with each contribution equal to the winding number of the complex phase angle of $\varepsilon_{\mathbf{k}\alpha\uparrow}+i\Delta_{\mathbf{k}}^{\alpha}$ along a closed path perpendicular to the $k_z=0$ plane.

\textit{\textcolor{black}{Applied to A$_{2}$Cr$_3$As$_3$.---}} The point group of the quasi-1D A$_{2}$Cr$_{3}$As$_{3}$ family is $D_{3h}$, which includes a $C_3$-rotation about the $z$-axis and a mirror reflection about the $xy$-plane. The low-energy band structure of A$_{2}$Cr$_{3}$As$_{3}$ can be well captured by a three-band tight-binding (TB) model $H_{\text{TB}}$ in the absence of SOC\cite{Wu}, with the three relevant orbitals to be the $3d_{z^2}$, $3d_{xy}$ and $3d_{x^2-y^2}$ respectively. From symmetry analysis which is given in the Appendix \ref{appendixB}, the leading SOC in this family takes the following on-site formula\cite{Wu},
 \begin{equation}
 H_{\text{SOC}}=i\lambda_{\text{so}}\sum_{\mathbf{k}\sigma}\sigma \left[c^{\dag}_{2\sigma}(\mathbf{k})c_{3\sigma}(\mathbf{k})-c^{\dag}_{3\sigma}(\mathbf{k})c_{2\sigma}(\mathbf{k})\right], \label{SOC}
\end{equation}
which possesses the SUS required. We adopt $\lambda_{\text{so}}\approx 10$ meV\cite{Wu} below. Note that the mirror-reflection symmetry forbids spin-flipping on-site SOC, because each such term as $c^{\dagger}_{\mathbf{i}\mu\sigma}c_{\mathbf{i}\nu\overline{\sigma }}g_{\mu\nu}$ would be changed to $c^{\dagger}_{\mathbf{i}\mu\sigma}c_{\mathbf{i}\nu\overline{\sigma }}\sigma\overline{\sigma }g_{\mu\nu}=-c^{\dagger}_{\mathbf{i}\mu\sigma}c_{\mathbf{i}\nu\overline{\sigma }}g_{\mu\nu}$ under this symmetry operation as shown in Appendix \ref{appendixB}.

\begin{figure}
\includegraphics[width=1.0\columnwidth]{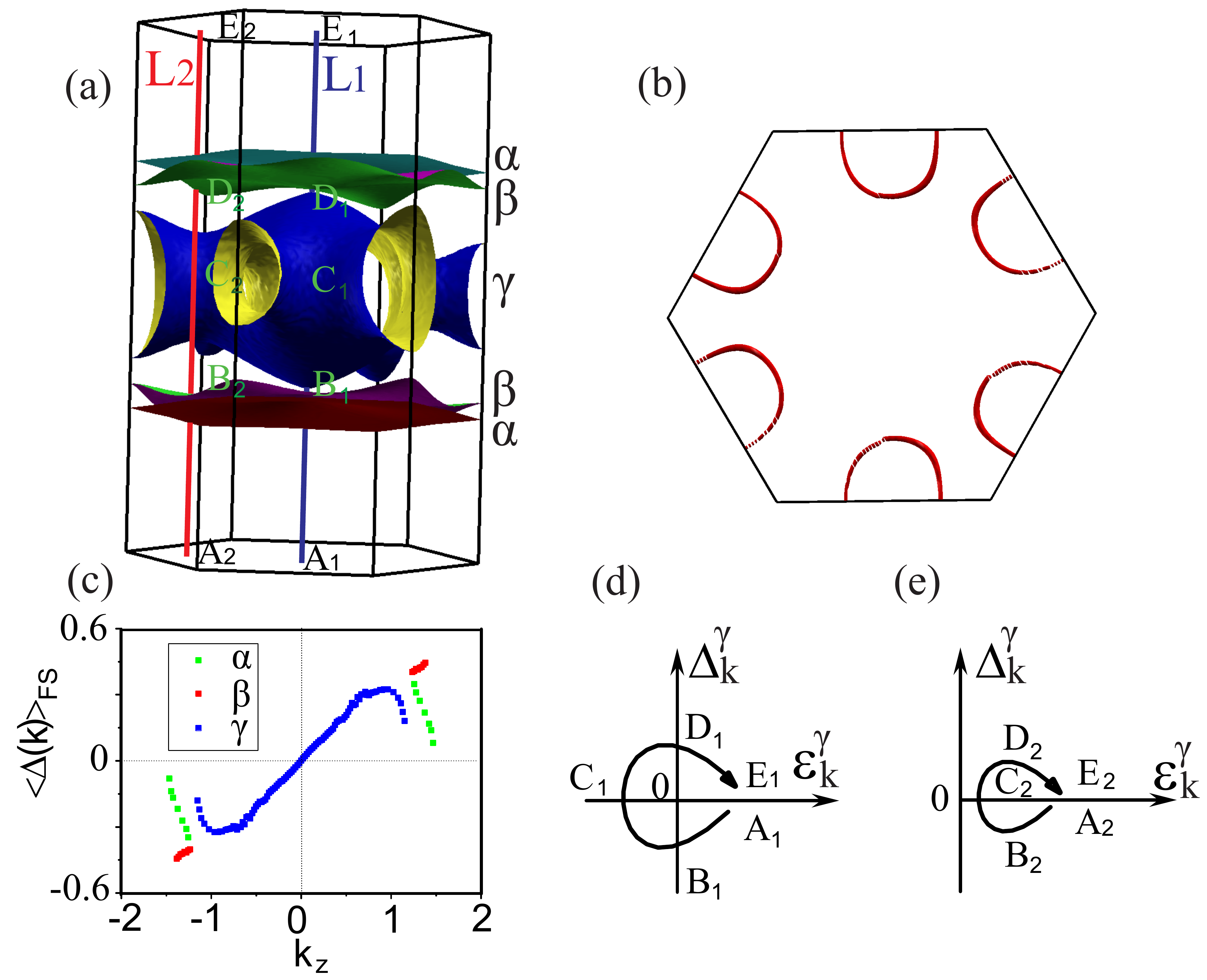}
\caption
{(Color online) FSs for the spin-up electrons and pairing gap function of K$_{2}$Cr$_{3}$As$_{3}$. (a) FSs of the TB model with SOC for K$_{2}$Cr$_{3}$As$_{3}$. The paths $L_{i,i=1,2}$ are perpendicular to the $(k_x,k_y)$-plane. (b) The intersection lines between the $\gamma$-FS and the $(k_x,k_y)$-plane, which are also the nodal lines of the $p_z$-SC. (c) The $k_z$-dependence of the relative gap function of K$_{2}$Cr$_{3}$As$_{3}$ averaged on the FSs.  Schematic diagrams of how the phase angle of $\varepsilon_{\mathbf{k}\gamma\uparrow}+i\Delta_{\mathbf{k}}^{\gamma}$  evolve with $k_z$ along the paths $L_1$ for (d) and $L_2$ for (e).}
\label{FS}
\end{figure}

The FSs of the spin-up electrons shown in Fig.~\ref{FS}(a) consists of two 1D FSs named as $\alpha$ and $\beta$ and one 3D FS named as $\gamma$. While each 1D FS contains two FS sheets nearly parallel to the $(k_x,k_y)$- plane, the 3D $\gamma$-FS intersects with the $(k_x,k_y)$- plane with their intersection line shown in Fig.~\ref{FS}(b). Note that the shape of the $\gamma$-FS is counter-intuitive: it contains one connected large concave pocket centering around the $\Gamma$-point, instead of three isolated small convex pockets centering around the $M$-points. The FSs of the spin-down electrons are related to those of the spin-up electrons through the relation $\varepsilon_{\mathbf{k}\alpha\downarrow}=\varepsilon_{-\mathbf{k}\alpha\uparrow}$ brought about by the TRS, and the lack of inversion symmetry leads to $\varepsilon_{-\mathbf{k}\alpha\uparrow}\ne\varepsilon_{\mathbf{k}\alpha\uparrow}$ and hence $\varepsilon_{\mathbf{k}\alpha\downarrow}\ne\varepsilon_{\mathbf{k}\alpha\uparrow}$, which means that the band structures of the two spin-species don't coincide.

Both weak-coupling RPA-based calculations and strong-coupling mean-field results suggest that the leading pairing symmetry of the system is TRI $p_z$-wave pairing with a dominating triplet component in the $S_z=0$ channel with line nodes\cite{Wu,Zhang} consistent with experiment\cite{Bao, Tang1,Tang2,uncsc1}, conserving the SUS. Therefore, the A$_2$Cr$_3$As$_3$ family are expected to belong to the symmetry class required here.  Since the $T_c$ ( $\le 9$ K ) of A$_{2}$Cr$_{3}$As$_{3}$ is much lower than the low-energy band width ($\approx 100$ meV), its pairing state can be well approximated as intra-band pairing, wherein Eq.(\ref{intraband_limit}) applies.

The gap function of the $p_z$-wave pairing obtained by the RPA approach is $C_3$-rotation invariant about the $z$-axis and doesn't obviously depend on $k_{x/y}$. The $k_z$-dependence of the relative gap function averaged on the FSs is shown in Fig.~\ref{FS}(c), where the sign of $\Delta^{\alpha}_{\mathbf{k}}(\alpha=1,2,3)$ follows that of $k_z$. Let's take the $\gamma$-band as an example to illustrate how to use Eq.(\ref{intraband_limit}) to calculate $I_{\gamma}$. Figure~\ref{FS}(d) and (e) illustrate in a schematic manner how the complex phase angle of $\varepsilon_{\mathbf{k}\gamma\uparrow}+i\Delta_{\mathbf{k}}^{\gamma}$ evolves from the A$_i$ to E$_i$ points along the two vertical lines $L_{i(i=1,2)}$ in the BZ shown in Fig.~\ref{FS}(a). Clearly, because the $L_1$ path passes the $\gamma$-FS twice which leads to twice sign changes of $\varepsilon_{\mathbf{k}\gamma\uparrow}$ and that the $p_z$-symmetry leads to sign change of $\Delta_{\mathbf{k}}^{\gamma}$ on the two $\gamma$-FS sheets, a nontrivial winding number $I_{\gamma}(L_1)=-1$ of the phase angle is obtained. On the contrary, the $L_2$ path doesn't pass the $\gamma$-FS, which leads to no sign change of $\varepsilon_{\mathbf{k}\gamma\uparrow}$ and hence $I_{\gamma}(L_2)=0$. Similarly, $I_{\alpha/\beta}(L_{1/2})=-1$. As a result $Z(k_x,k_y)|_{L_1}=-3$ and $Z(k_x,k_y)|_{L_2}=-2$. Therefore, the three elliptical areas centered around the M points (the remaining part) in Fig.~\ref{FS}(b) are covered by $Z=-2$ ($-3$), which will lead to 2 (3) flat bands over this regime on the surface BZ on each (001) surface.

Note that the topological properties obtained here are protected by the SUS. Supposing a vanishingly weak spin-flipping SOC turns on, the system now belongs to conventional TRI superconductors, whose topological invariant $Z$ is then defined for the whole off-diagonal block $q$ of Eq.(\ref{Q_matrix})\cite{Schnyder,Chiu2016,Schnyder2011} which satisfies $Z(k_x,k_y)=Z_1(k_x,k_y)+Z_2(k_x,k_y)=Z_1(k_x,k_y)+Z_1(-k_x,-k_y)$. In the case with weak on-site SOC with SUS for the $p_z$-wave SC, we have $\varepsilon_{\mathbf{k}\gamma\uparrow}=\varepsilon_{\mathbf{-k}\gamma\downarrow}\approx \varepsilon_{\mathbf{-k}\gamma\uparrow}; \Delta_{\mathbf{k}}^{\gamma}=-\Delta_{\mathbf{-k}}^{\gamma}$, leading to $\theta_{\mathbf{k}\alpha}\approx-\theta_{\mathbf{-k}\alpha}$. Then from Eq.(\ref{intraband_limit}) we find that except in a narrow regime to be studied below, in most regime of the $(k_x,k_y)$- plane we have $Z_1(k_x,k_y)=-Z_1(-k_x,-k_y)$ for integer $Z_1$, and hence $Z(k_x,k_y)=0$. Here the protection of the SUS permits that we only count $Z_1$, which is nonzero all over the $(k_x,k_y)$- plane.
\begin{figure}
\includegraphics[width=1.0\columnwidth]{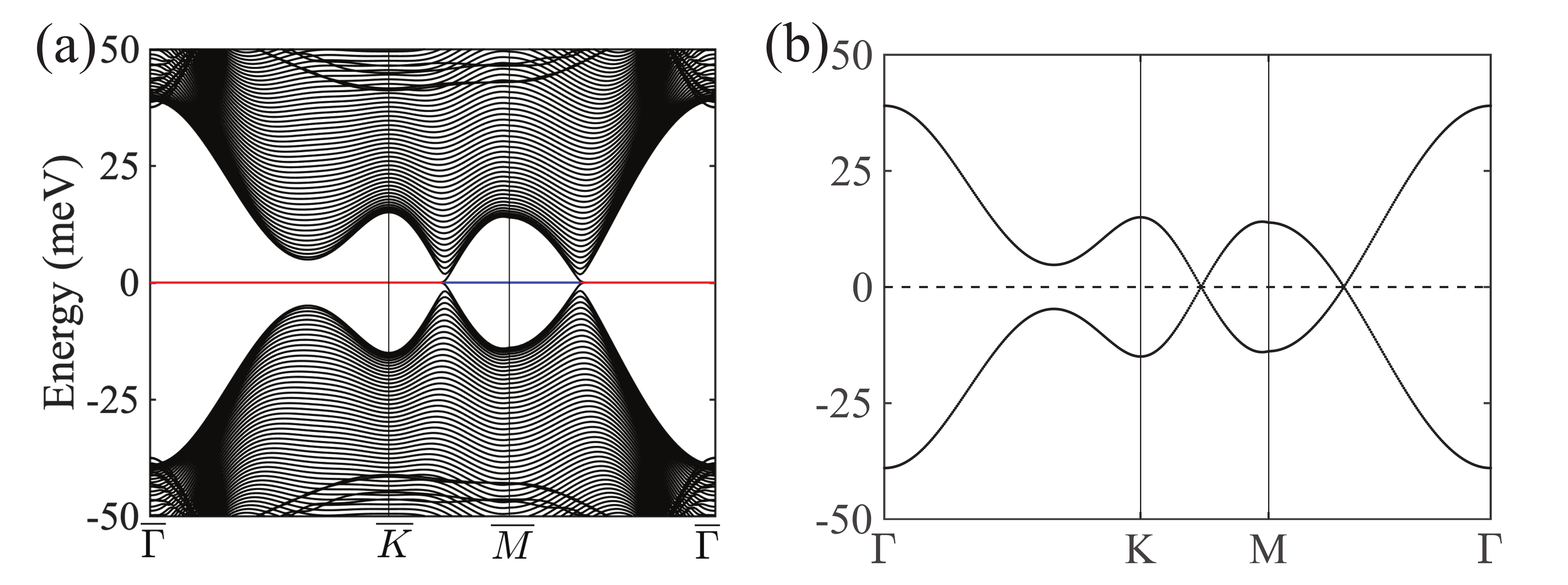}
\caption
{(Color online) (a) The energy spectrum  as function of $k_{x/y}$ with open boundary condition along the $z$-axis and periodic ones along the $x$- and $y$- axes. (b) The bulk bands in the $p_z$-wave pairing state with fixed $k_z=0$. In (a), the segment marked red (blue) is covered by 6 (4) flat bands.  The number of the slab layers is 200. The adopted $\Delta_{1}$=20 meV, $\Delta_{2}=\Delta_{3}$=40 meV are enhanced by an order of magnitude over realistic ones to enhance the visibility.}
\label{SS}
\end{figure}

\textit{\textcolor{black}{Surface spectrum and STM.---}} The nontrivial topological invariant of A$_{2}$Cr$_{3}$As$_{3}$ leads to flat surface bands. We have studied the edge spectrum of this system on the (001) surface as shown in Appendix \ref{appendixC}. The obtained energy spectrum as function of $k_x$ and $k_y$ is shown in Fig.~\ref{SS}(a) along the high symmetric line, in comparison with the bulk band in the superconducting state shown in Fig.~\ref{SS}(b) with fixed $k_z=0$. To enhance visibility, the adopted pairing gap amplitudes are enhanced from realistic $\Delta_{1}\approx$ 1 meV, $\Delta_{2}=\Delta_{3}\approx$ 2 meV\cite{Wu} for K$_2$Cr$_3$As$_3$ with $T_c\approx 6$ K to $\Delta_{1}$=20 meV, $\Delta_{2}=\Delta_{3}$=40 meV. The comparison between Fig.~\ref{SS}(a) and (b) suggests that, in addition to the bulk continuum, some regime in the $(k_x,k_y)$-plane is covered by extra 6 flat bands while the remaining regime is covered by extra 4, with the boundary of the two regimes to be just the SC nodal lines shown in Fig.~\ref{FS}(b).

\begin{figure}
\includegraphics[width=1.0\columnwidth]{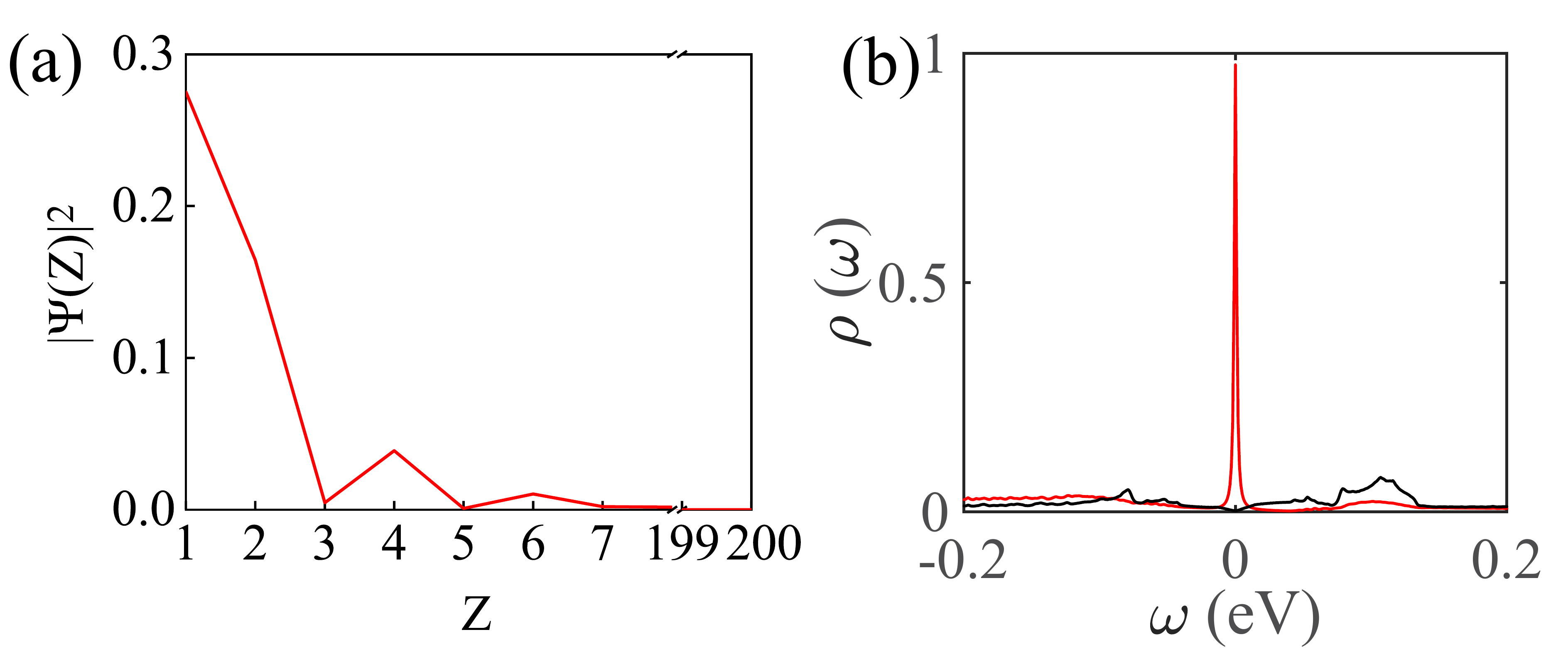}
\caption
{(Color online) (a) Distribution of the squared modulus of the wave function of the particle part of the Bogolubov quasi-particle (the hole-part is similar) along the $z$ direction for a typical state in the flat bands. (b) The differential conductance $dI/dV\sim V$ spectra of the STM for the end (red) and the middle (black) of K$_2$Cr$_3$As$_3$.}
\label{STM}
\end{figure}

The flat bands shown in Fig.~\ref{SS}(a) are formed by bound states localized at the two (001) surfaces, which is justified by the distribution of the wave functions of the Bogolubov quasi-particles shown in Fig.~\ref{STM}(a), which illustrates a bound state with a localized length $\xi\approx 3c$ with the lattice constant $c=4.23$ $\AA$. Our numerical results suggest $\xi\approx 45c\approx20$ nm for realistic gap amplitudes. As the two (001) surfaces symmetricly share the surface states, the corresponding areas in Fig.~\ref{SS}(a) are covered by 3 (2) flat bands on each surface BZ, consistent with the topological invariant calculations above.

The topological flat surface bands obtained above can be detected as the ZBCP in the site-dependent differential conductance spectrum $dI/dV$ of the STM. Details for the calculation of $dI/dV$ are provided in the Appendix \ref{appendixD}. Figure \ref{STM}(b) shows the $dI/dV\sim V$ curves for the end (red) and middle (black) points of the needle-like sample of the quasi-1D material respectively. Obviously, there is a very sharp ZBCP in the spectrum of the end point, which is absent in that of the middle point.

\begin{figure}
\includegraphics[width=1.0\columnwidth]{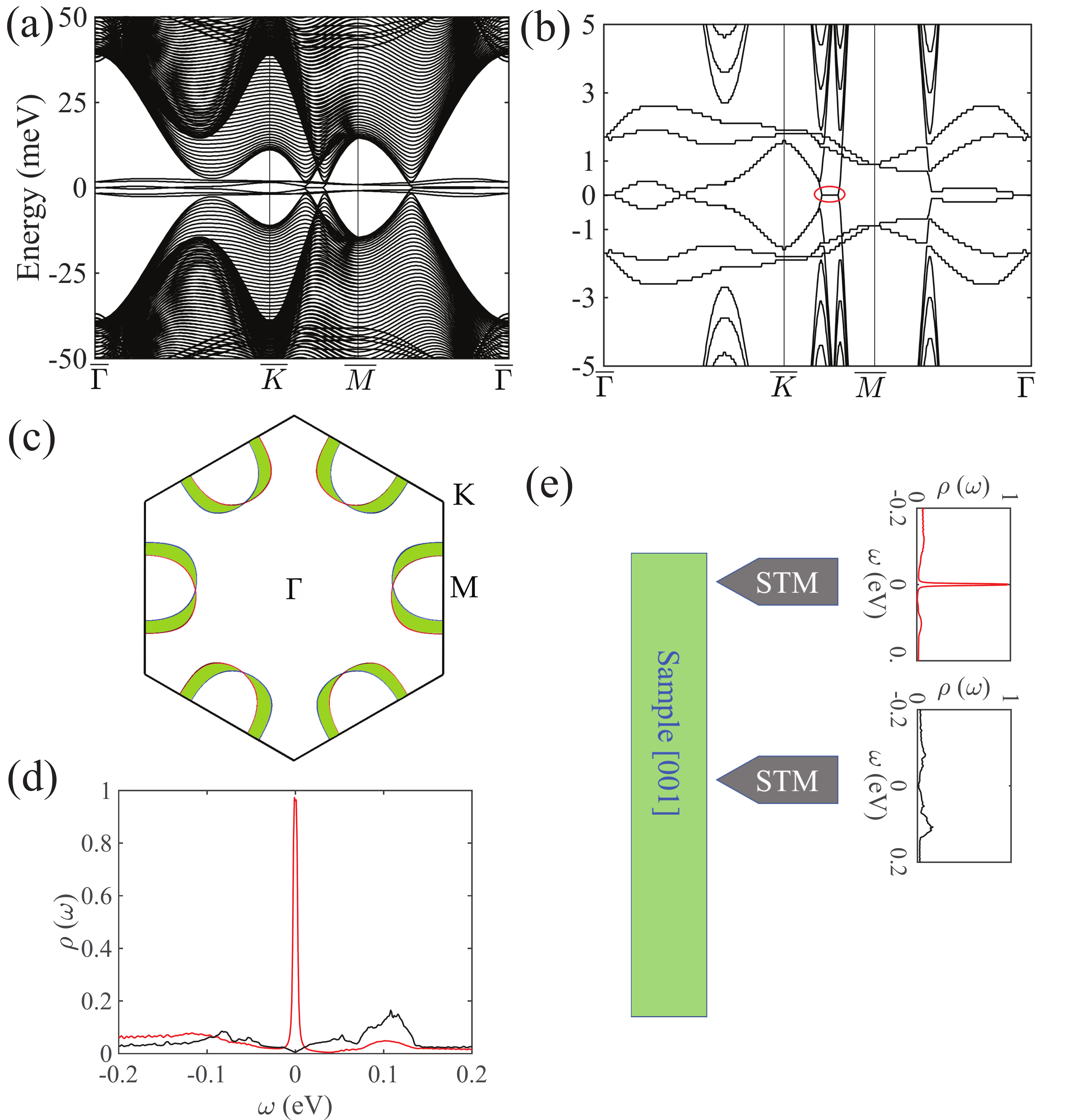}
\caption
{(Color online) The surface spectrum (a) and its zoom-in (b) with NN-spin-flipping SOC with strength $\lambda_{1}=\lambda_{2}=2$ meV for K$_{2}$Cr$_{3}$As$_{3}$. (c) The noncentrosymmetric distribution of $Z_1(k_x,k_y)$ in the surface BZ, with the narrow shaded regime covered by $|Z|=1$. (d) The STM spectra for the end (red) and the middle (black) of the sample respectively. (e) Schematic configuration for experimental identification of the $p_z$-wave SC in the quasi-1D A$_{2}$Cr$_{3}$As$_{3}$ family.}
\label{TSC}
\end{figure}

\textit{\textcolor{black}{Spin-flipping SOC.---}} In real material of A$_{2}$Cr$_{3}$As$_{3}$, there can be weak subleading NN-spin-flipping SOC, whose explicit formula is given in Appendix \ref{appendixB}. This SOC term breaks the SUS, and the topological invariant Eq.(\ref{windingnumber}) doesn't apply. However, as this NN-SOC for the $3d$-orbitals is so weak (with strength $\lambda_{1/2}=2$ meV adopted) that the above obtained flat bands are only slightly split, as shown in Fig.~\ref{TSC}(a) and its zoom-in in Fig.~\ref{TSC}(b). As a result, the sharp ZBCP is still present in the STM spectrum shown in Fig.~\ref{TSC}(d).

Remarkably, even the spin-flipping SOC breaks the SUS here, there is still a narrow but finite regime in the BZ covered by a pair of exactly-flat bands, as highlighted by the red oval in Fig.~\ref{TSC}(b). This pair of exactly-flat bands are protected by the topological invariant for conventional TRI SC without SUS\cite{Schnyder,Chiu2016,Schnyder2011}. As introduced above, for sufficiently weak $\lambda_{1/2}$, $Z(k_x,k_y)=Z_1(k_x,k_y)+Z_1(-k_x,-k_y)$. As the A$_{2}$Cr$_{3}$As$_{3}$ family is noncentrosymmetric, the weak difference between $\varepsilon_{\mathbf{k}\alpha\uparrow}$ and $\varepsilon_{\mathbf{-k}\alpha\uparrow}$ caused by Eq.(\ref{SOC}) leads to weak noncentrosymmetry in $|Z_1(k_x,k_y)|$ and hence nonzero $Z(k_x,k_y)=\pm 1$ in the narrow shaded regime in Fig.~\ref{TSC}(c), causing a pair of exactly-flat surface bands there.

\textit{\textcolor{black}{Discussion and Conclusion.---}} One may worry that the bad quality of the surface and the breaking of mirror-reflection symmetry there might hinder the detection of the surface states. The solution of these problems is shown in Fig.~\ref{TSC}(e): when the tip of the STM is put on the side surface near the end of the sample within the localized length $\xi\approx45c\approx20$, there would be pronounced ZBCP in the spectrum, and when the tip is far from the end the ZBCP would vanish. Note that we can use the STM configuration adopted recently\cite{liu2019} to distinguish between the edge spectra of the A$_{2}$Cr$_{3}$As$_{3}$ and ACr$_{3}$As$_{3}$: while the former would exhibit the ZBCP,  the latter with $s^{\pm}$- pairing\cite{133} would not. Interestingly, the number of surface flat bands here can be easily tuned via doping as shown in Appendix \ref{appendixE}, readily tested by experiments.

The SUS-protected topological invariant proposed here also applies to other unconventional superconductors with weak SOC, particularly those with mirror-reflection symmetry is proved in Appendix \ref{appendixB}, which maintains the SUS required here.

In conclusion, we have proposed an SUS protected momentum-dependent integer-$Z$-valued topological invariant for TRI superconductors, whose nonzero value will lead to exactly-flat surface bands. The projection of the bulk nodal line onto the surface BZ serves as a boundary across which the topological invariant changes $\pm 1$, which can be nonzero on both sides due to their integer-Z-valued character, distinguished from conventional topological nodal-line superconductors. Applying this theory to the A$_{2}$Cr$_{3}$As$_{3}$ family up to the leading SOC, we find that the whole (001) surface BZ is covered with exactly-flat bands, which can be detected by the STM as sharp ZBCP. Probably, the band-flattening on the surface might drive new instabilities to be detected. Our discovery not only reveals a new type of TRI TSC, but it also provides smoking gun evidence for the experimental identification of the $p_z$-wave pairing symmetry of the quasi-1D A$_{2}$Cr$_{3}$As$_{3}$ family.

\section*{Acknowledgements}
We are grateful to the helpful discussions with Guo-Qing Zheng, Hong-Ming Weng, Jiangping Hu and Yugui Yao. This work is supported by the NSFC (Grant Nos. 11729402,11922401,11674025, 11774028,11604013), and Basic Research Funds of Beijing Institute of Technology (No. 2017CX01018). Cheng-Cheng Liu and Chen Lu contributed equally to this work.

\appendix

\section{Topological Invariant}\label{appendixA}

In this section, we derive the spin-U(1)-symmetry (SUS) protected momentum-dependent integer-Z-valued topological invariant for time-reversal-invariant (TRI) superconductors, whose nonzero value will lead to exactly flat band(s) on the surface Brillouin zone.

We start from the following $N_\alpha$-band model:
\begin{align}\label{H1}
H=\sum_{\mathbf{k}\alpha\sigma } C_{\mathbf{k}\alpha \sigma }^{\dagger}C_{\mathbf{k}\alpha \sigma } \varepsilon_{\mathbf{k}\sigma}^{\alpha }+\sum_{\mathbf{k}\alpha\beta} \left[C_{\mathbf{k}\alpha \uparrow }^{\dagger}C_{\mathbf{-k}\beta \downarrow }^{\dagger}\Delta_{\alpha \beta }(\mathbf{k})+h.c.\right].
\end{align}
Here $\alpha/\beta=1,\cdots,N_\alpha$ denote the band indices. Note that here we have allowed SOC with SUS and inter-band pairing. From time-reversal symmetry (TRS), we obtain $\varepsilon_{\mathbf{k}\uparrow}^{\alpha }=\varepsilon_{\mathbf{-k}\downarrow}^{\alpha }$ and $\Delta(\mathbf{k})=\Delta^{\dagger}(\mathbf{k})$. We can rewrite $H$ into the formula of
\begin{align}\label{H2}
H=\frac{1}{2}\sum_{\mathbf{k}}\begin{pmatrix}
C_{\mathbf{k} \uparrow }^{\dagger}& C_{\mathbf{k} \downarrow }^{\dagger} & C_{\mathbf{-k} \uparrow }&C_{\mathbf{-k} \downarrow }
\end{pmatrix}    \begin{pmatrix}
H_{\mathbf{k}}
\end{pmatrix}\begin{pmatrix}
C_{\mathbf{k} \uparrow }\\
C_{\mathbf{k} \downarrow }\\
C_{\mathbf{-k} \uparrow }^{\dagger}\\
C_{\mathbf{-k} \downarrow }^{\dagger}
\end{pmatrix},
\end{align}
with $C_{\mathbf{k} \uparrow }^{\dagger}\equiv (\cdots C_{\mathbf{k}\alpha \uparrow }^{\dagger}\cdots)$, and \begin{align}\label{H21}
H_{\mathbf{k}}=\begin{pmatrix}
\varepsilon_{\mathbf{k}\uparrow}& 0 & 0 & \Delta_{\mathbf{k}}\\
0& \varepsilon_{\mathbf{k}\downarrow} & -\Delta^{T}_{\mathbf{-k}} & 0\\
0& -\Delta^{T}_{\mathbf{-k}} & -\varepsilon_{\mathbf{-k}\uparrow} & 0\\
\Delta_{\mathbf{k}}& 0 & 0 & -\varepsilon_{\mathbf{-k}\downarrow}
\end{pmatrix}
\end{align} is a $4N_{\alpha}\times 4N_{\alpha}$ matrix. Here $\varepsilon_{\mathbf{k}\sigma}\equiv \text{diag}(\cdots \varepsilon_{\mathbf{k}\alpha\sigma}\cdots)$.
Let's perform the following unitary transformation on $H_{\mathbf{k}}$

\begin{align}\label{H3}
H_{\mathbf{k}}\rightarrow \widetilde{H}_{\mathbf{k}}=U^{\dagger}HU,
\end{align}
where $U=\frac{1}{\sqrt{2}}\begin{pmatrix}
I & I\\
\sigma _y& -\sigma _y
\end{pmatrix}$ and
\begin{align}\label{H4}
\widetilde{H}_{\mathbf{k}}= \left(\begin{array}{cccc}
0&0&\varepsilon_{\mathbf{k}\uparrow}-i\Delta_{\mathbf{k}}&0 \\
0 & 0& 0&\varepsilon_{\mathbf{k}\downarrow}-i\Delta^{T}_{\mathbf{-k}} \\
\varepsilon_{\mathbf{k}\uparrow}+i\Delta_{\mathbf{k}}& 0&0&0\\
0 & \varepsilon_{\mathbf{k}\downarrow}+i\Delta^{T}_{\mathbf{-k}}&0 & 0\\
\end{array}\right).
\end{align}
Note that
\begin{align}\label{H5}
\widetilde{H}_{\mathbf{k}}=\left(\begin{array}{cccc}
1& 0 &  0&0 \\
0 & 0 &  1& 0\\
0&  1& 0 &0 \\
0 & 0 & 0 &1
\end{array}\right)^{\dagger} \widetilde{\widetilde{H}}_{\mathbf{k}}\left(\begin{array}{cccc}
1& 0 &  0&0 \\
0 & 0 &  1& 0\\
0&  1& 0 &0 \\
0 & 0 & 0 &1
\end{array}\right),
\end{align}
with \begin{align}\label{H51}\widetilde{\widetilde{H}}_{\mathbf{k}}= \left(\begin{array}{cccc}
0& \varepsilon_{\mathbf{k}\uparrow}-i\Delta_{\mathbf{k}} & 0 & 0\\
\varepsilon_{\mathbf{k}\uparrow}+i\Delta_{\mathbf{k}}& 0 & 0 & 0\\
0& 0 & 0 & \varepsilon_{\mathbf{k}\downarrow}-i\Delta^{T}_{\mathbf{-k}}\\
0& 0& \varepsilon_{\mathbf{k}\downarrow}+i\Delta^{T}_{\mathbf{-k}}& 0\\
\end{array}\right).\end{align}

Let's study the eigenvalues and eigenvectors of $\widetilde{\widetilde{H}}_{\mathbf{k}}$ and $\widetilde{H}_{\mathbf{k}}$. It's proved here that for the fully-gapped case of $\widetilde{\widetilde{H}}_{\mathbf{k}}$, the 1-2 (3-4) diagonal block contributes $N_{\alpha}$ negative eigenvalues and $N_{\alpha}$ positive ones respectively, with the corresponding eigenvectors in the form of $\begin{pmatrix}
\mu^{T},&
\nu^{T},&
0,&
0
\end{pmatrix}^{T}$ and $\begin{pmatrix}
\mu^{T},&
-\nu^{T},&
0,&
0
\end{pmatrix}^{T}$ ($\begin{pmatrix}
0,&
0,&
\mu^{T},&
\nu^{T}
\end{pmatrix}^{T}$ and $\begin{pmatrix}
0,&
0,&
\mu^{T},&
-\nu^{T}
\end{pmatrix}^{T}$). Here $\mu$ and $\nu$ are $N_{\alpha}$-component column vector. Actually, to solve the eigenvalue problem of the 1-2 diagonal block of $\widetilde{\widetilde{H}}_{\mathbf{k}}$ in Eq.(\ref{H51}), one needs to solve the equation
\begin{align}\label{H60}
\det \left [
\left(\begin{array}{cc}
0 &\varepsilon_{\mathbf{k}\uparrow}-i\Delta_{\mathbf{k}} \\
\varepsilon_{\mathbf{k}\uparrow}+i\Delta_{\mathbf{k}} &0 \\
\end{array}\right)-\lambda I
\right ]=0
\end{align}
with $\lambda$ to be the eigenvalue. Since
\begin{align}\label{H6}
&\det \left [
\left(\begin{array}{cc}
0 &\varepsilon_{\mathbf{k}\uparrow}-i\Delta_{\mathbf{k}} \\
\varepsilon_{\mathbf{k}\uparrow}+i\Delta_{\mathbf{k}} &0 \\
\end{array}\right)-\lambda I
\right ]  \nonumber\\
&=
\det
\left(\begin{array}{cc}
-\lambda I &\varepsilon_{\mathbf{k}\uparrow}-i\Delta_{\mathbf{k}} \\
\varepsilon_{\mathbf{k}\uparrow}+i\Delta_{\mathbf{k}} &-\lambda I
\end{array}\right)\equiv\det
\left(\begin{array}{cc}
-\lambda I &F \\
F^{\dagger} &-\lambda I
\end{array}\right)   \nonumber\\
&=\det
\left(\begin{array}{cc}
-\lambda I &F \\
F^{\dagger} &-\lambda I
\end{array}\right)\cdot \det
\left(\begin{array}{cc}
I &\lambda^{-1} F \\
0 &I
\end{array}\right) \nonumber\\
&=\det
\left(\begin{array}{cc}
-I &0 \\
F^{\dagger} &F^{\dagger}F-\lambda^{2} I
\end{array}\right),
\end{align}
we have $\det(F^{\dagger}F-\lambda^{2} I)=0$. Because $F^{\dagger}F$ is a Hermitian operator with positive-definite eigenvalues, the obtained values of $\lambda$ are always positive-negative symmetrically distributed. Therefore, the 1-2 diagonal block of the $\widetilde{\widetilde{H}}_{\mathbf{k}}$ in Eq.(\ref{H51}) contributes $N_{\alpha}$ negative eigenvalues and $N_{\alpha}$ positive ones respectively, with the corresponding eigenvectors in the form of $\begin{pmatrix}\mu^{T},&\nu^{T},&0,&0\end{pmatrix}^{T}$ and $\begin{pmatrix}\mu^{T},&-\nu^{T},&0,&0\end{pmatrix}^{T}$. Similarly, the 3-4 diagonal block of the $\widetilde{\widetilde{H}}_{\mathbf{k}}$ in Eq.(\ref{H51}) contributes the same number of negative and positive eigenvalues, with the corresponding eigenvectors in the form of $\begin{pmatrix}0,&0, &\mu^{T},&\nu^{T}\end{pmatrix}^{T}$ and $\begin{pmatrix}0,&0, &\mu^{T},&-\nu^{T}\end{pmatrix}^{T}$. Then from Eq.(\ref{H5}), the distribution of the eigenvalues of $\tilde{H}_{\mathbf{k}}$ is similar, but with corresponding eigenvectors in the form of $\begin{pmatrix}\mu^{T},&0,&\nu^{T},&0\end{pmatrix}^{T}$ and $\begin{pmatrix}\mu^{T},&0,&-\nu^{T},&0\end{pmatrix}^{T}$ ($\begin{pmatrix}0,&\mu^{T},&0,&\nu^{T}\end{pmatrix}^{T}$ and $\begin{pmatrix}0,&\mu^{T},&0,&-\nu^{T}\end{pmatrix}^{T}$).

Following the standard procedure for the topological invariant of TRI SC\cite{Schnyder, Chiu2016}, we evaluate the projection operator $\widehat{P}$ as
\begin{align}\label{H7}
\widehat{P}=&\sum_{i=1}^{N_{\alpha}}\begin{pmatrix}
\mu_{i}\\
0\\
\nu_{i}\\
0
\end{pmatrix}\times \begin{pmatrix}
\mu_{i}^{\dagger} & 0 &\nu_{i}^{\dagger}   & 0
\end{pmatrix}+\begin{pmatrix}
0\\
\mu^{\prime}_{i}\\
0\\
\nu^{\prime}_{i}
\end{pmatrix}\times \begin{pmatrix}
0 & \mu^{\prime\dagger}_{i} &0   & \nu^{\prime\dagger}_{i}
\end{pmatrix} \nonumber\\
=&\begin{pmatrix}
\frac{I}{2} & 0 & -\frac{q_1}{2} & 0\\
0&\frac{I}{2}  & 0 &-\frac{q_2}{2} \\
-\frac{q^{\dagger}_1}{2}& 0 & \frac{I}{2} & 0\\
0 & -\frac{q^{\dagger}_2}{2} & 0 &\frac{I}{2}
\end{pmatrix}.
\end{align}
Here $\begin{pmatrix}
\mu_{i}\\
0\\
\nu_{i}\\
0\end{pmatrix}$ and $\begin{pmatrix}
0\\ \mu^{\prime}_{i}\\
0\\
\nu^{\prime}_{i}\end{pmatrix}$ ($i=1,\cdots,N_{\alpha}$) are the $2N_{\alpha}$ eigenvectors of Eq.(\ref{H4}) with negative eigenvalues:
\begin{eqnarray}\label{eigen}
\widetilde{H}_{\mathbf{k}}\begin{pmatrix}
\mu_{i}\\
0\\
\nu_{i}\\
0
\end{pmatrix}=E_i(\mathbf{k})\begin{pmatrix}
\mu_{i}\\
0\\
\nu_{i}\\
0
\end{pmatrix},E_i(\mathbf{k})<0\nonumber\\
\widetilde{H}_{\mathbf{k}}\begin{pmatrix}
0\\\mu_{i}^{\prime}\\
0\\\nu_{i}^{\prime}\end{pmatrix}=E^{\prime}_i(\mathbf{k})\begin{pmatrix}
0\\\mu_{i}^{\prime}\\
0\\\nu_{i}^{\prime}\end{pmatrix},E^{\prime}_i(\mathbf{k})<0.
\end{eqnarray}
The formulae of $q_1,q_2$ are
\begin{eqnarray}\label{q12}
q_1&=&-2\sum_{i=1}^{N_{\alpha}}\mu_{i}\nu_{i}^{\dagger}\nonumber\\
q_2&=&-2\sum_{i=1}^{N_{\alpha}}\mu^{\prime}_{i}\nu^{\prime\dagger}_{i}
\end{eqnarray}
Here we have used the relation $\sum_{i=1}^{N_{\alpha}}\mu_{i}\mu_{i}^{\dagger}+\sum_{i=1}^{N_{\alpha}}\mu_{i}\mu_{i}^{\dagger}=I\to\sum_{i=1}^{N_{\alpha}}\mu_{i}\mu_{i}^{\dagger}=I/2; \sum_{i=1}^{N_{\alpha}}\nu_{i}\nu_{i}^{\dagger}+\sum_{i=1}^{N_{\alpha}}(-\nu_{i})(-\nu_{i}^{\dagger})=I\to\sum_{i=1}^{N_{\alpha}}\nu_{i}\nu_{i}^{\dagger}=I/2$ and similarly $\sum_{i=1}^{N_{\alpha}}\mu_{i}^{\prime}\mu_{i}^{\prime\dagger}=I/2; \sum_{i=1}^{N_{\alpha}}\nu_{i}^{\prime}\nu_{i}^{\prime\dagger}=I/2$. Then the $\widehat{Q}$ operator is
\begin{align}\label{H8}
\widehat{Q}=1-2\widehat{P}=\left(\begin{array}{cccc}
0&0&q_1&0\\
0&0&0&q_2\\
q^{\dagger}_1&0&0&0\\
0&q^{\dagger}_2&0&0\\
\end{array}\right).
\end{align}
Note that $\widehat{Q}$ is $2\times2$ block-off-diagonal, and furthermore the off-diagonal block is block-diagonal. Due to this property, the winding number can be defined as $Z_{1/2}$, with \begin{equation}\label{Topological_invariant}
Z_{1/2}(k_x,k_y)=\frac{i}{2\pi }\int_{-\pi }^{\pi } tr(q^{\dagger}_{1/2}\partial_{k_z} q_{1/2})dk_z.
\end{equation}
Note that $Z_1(k_x,k_y)\left[\left\{\varepsilon_{\mathbf{k}},\Delta_{\mathbf{k}}\right\}\right]$$=$$Z_2(-k_x,-k_y)\left[\left\{\varepsilon_{\mathbf{k}},\Delta^{T}_{\mathbf{k}}\right\}\right]$.

For nontrivial $Z_{1/2}(k_x,k_y)\ne 0$, there will be $|Z_1|+|Z_2|$ zero-modes at the momentum $(k_x,k_y)$ on the surface Brillouin zone in the above 4-component Nambu representation (\ref{H2}). However, the gauge redundancy in this representation brings up double counting on the number of zero-modes for each momentum. In fact, due to the SUS here, one can also write the BdG Hamiltonian in the gauge-redundancy-free 2-component Nambu-representation as $H=\sum_{\mathbf{k}}\psi_{\mathbf{k}}^{\dagger}h\psi_{\mathbf{k}}$, with $\psi^{\dagger}_{\mathbf{k}}=\left(\mathbf{c_{\mathbf{k}\uparrow}^{\dagger}},\mathbf{c_{-\mathbf{k}\downarrow}}\right)$, and $h$ is equal to the 1-4 block of Eq.(\ref{H21}). In this  representation, the number of zero-modes at momentum $(k_x,k_y)$ is $|Z_1(k_x,k_y)|$, and the other $|Z_2(k_x,k_y)|$ zero-modes obtained in the 4-component representation are just those folded from the momentum $(-k_x,-k_y)$ artificially in that representation, which is a double counting. Therefore, the physical topological invariant here is $Z_1(k_x,k_y)$, which leads to $|Z_1(k_x,k_y)|$ flat surface bands.

In the following, we consider the special case of intra-band pairing limit, which is the case of most of the existing superconductors. Let's set $\Delta ^{\alpha \beta }(\mathbf{k})= \Delta ^{\alpha}(\mathbf{k})\delta _{\alpha \beta }$ and perform the calculations provided by Eq.(\ref{H4}), Eq.(\ref{eigen}) and Eq.(\ref{q12}). As a result, we obtain
\begin{eqnarray}
q_1=\text{diag}(e^{-i\theta^{1}_{\mathbf{k}1}} ,e^{-i\theta^{2}_{\mathbf{k}1}},\cdots,e^{-i\theta^{N_{\alpha }}_{\mathbf{k}1}})\nonumber\\
q_2=\text{diag}(e^{-i\theta^{1}_{\mathbf{k}2}} ,e^{-i\theta^{2}_{\mathbf{k}2}},\cdots,e^{-i\theta^{N_{\alpha }}_{\mathbf{k}2}})
\end{eqnarray}
with
\begin{eqnarray}
e^{-i\theta^{\alpha}_{\mathbf{k}1}}&=&(\varepsilon^{\alpha }_{\mathbf{k}\uparrow}-i\Delta^{\alpha }_{\mathbf{k}})/|\varepsilon^{\alpha }_{\mathbf{k}\uparrow}-i\Delta^{\alpha }_{\mathbf{k}}|\nonumber\\
e^{-i\theta^{\alpha}_{\mathbf{k}2}}&=&(\varepsilon^{\alpha }_{\mathbf{k}\downarrow}-i\Delta^{\alpha }_{\mathbf{-k}})/|\varepsilon^{\alpha }_{\mathbf{k}\downarrow}-i\Delta^{\alpha }_{\mathbf{-k}}|
\end{eqnarray}
Then from Eq.(\ref{Topological_invariant}), we evaluate the topological invariant $Z_{1/2}(k_x,k_y)$ as
\begin{align}\label{H13}
Z_{1/2}(k_x,k_y)=&\frac{i}{2\pi }\int_{-\pi }^{\pi } tr(q^{\dagger}_{1/2}\partial_{k_z} q_{1/2})dk_z \nonumber\\
=&\frac{i}{2\pi }\sum_{\alpha=1}^{N_{\alpha}}  \int_{-\pi }^{\pi }  (e^{i\theta^{\alpha}_{\mathbf{k}1/2}}\partial_{k_z} e^{-i\theta^{\alpha}_{\mathbf{k}1/2}})dk_z\nonumber\\
=&\frac{1}{2\pi }\sum_{\alpha=1}^{N_{\alpha}}\int_{-\pi }^{\pi }  \partial_{k_z}\theta^{\alpha}_{\mathbf{k}1/2}dk_z \nonumber\\
=&\sum_{\alpha=1}^{N_{\alpha}}\frac{1}{2\pi }\int_{-\pi }^{\pi }  d\theta^{\alpha}_{\mathbf{k}1/2} \equiv\sum_{\alpha=1}^{N_{\alpha}}I_{\alpha}.
\end{align}
Eq.(\ref{H13}) suggests that in the intra-band pairing limit, $Z_{1/2}(k_x,k_y)$ is a summation of the contributions $I_{\alpha}$ from each band $\alpha$, with each contribution $I_{\alpha}$ equal to the winding number of the complex phase angle of $\varepsilon_{\mathbf{k}\alpha\sigma}+i\Delta_{\mathbf{\pm k}}^{\alpha}$ along a closed path perpendicular to the $k_z=0$ plane.


\section{Formula of SOC in the A$_2$Cr$_3$As$_3$ family}\label{appendixB}
The A$_2$Cr$_3$As$_3$ family are quasi-1D superconductors consisting of alkali-metal-atom-separated [(Cr$_3$As$_3$)$^{2-}$]$_{\infty}$ double-walled subnanotubes extending along the easy axis, defined as the z-axis here. The point group of the material is $D_{3h}$, which includes a $C_3$-rotation about the $z$-axis and a mirror reflection $M$ about the $xy$-plane. The low-energy degrees of freedom of the material are the Cr-3d orbitals, which include the d$_{z^2}$ (orbital 1), the d$_{xy}$ (orbital 2) and the d$_{x^2-y^2}$ (orbital 3).

Due to the $D_{3h}$ point group and the time-reversal-symmetry (TRS), our system with the three low energy d-orbitals has the following symmetry operators.

(1) The time-reversal operator $\widehat{T}$:
\begin{equation}\label{TRS}
\widehat{T}:\quad C_{\mathbf{i}\mu \uparrow }\rightarrow C_{\mathbf{i}\mu \downarrow}\quad , \quad C_{\mathbf{i}\mu \downarrow} \rightarrow -C_{\mathbf{i}\mu \uparrow }
\end{equation}

(2) The mirror-reflection operator $\widehat{M}$:
\begin{equation}\label{MR}
\widehat{M}:\quad C_{\mathbf{i}\mu \uparrow }\rightarrow C_{\mathbf{i^{'}}\mu \uparrow}\quad , \quad C_{\mathbf{i}\mu \downarrow} \rightarrow -C_{\mathbf{i^{'}} \mu \downarrow }
\end{equation}
with
\begin{equation}
\mathbf{i^{'}}=\widehat{M} \mathbf{i}
\end{equation}

(3) The 120$^o$-degree rotation $\widehat{C}_3^1$:
\begin{equation}\label{RT1}
\widehat{C}_3^1:\quad C_{\mathbf{i}\mu\sigma}\rightarrow C_{\mathbf{i}\nu\sigma}D^{(1)}_{\mu\nu}e^{-i\sigma\pi/3},
\end{equation}
with
\begin{eqnarray}\label{asd1}
D^{(1)}_{\mu\nu}=\begin{pmatrix}
1&0&0\\
0&\frac{-1}{2} & \frac{\sqrt{3}}{2}\\
0&\frac{-\sqrt{3}}{2}& \frac{-1}{2}
\end{pmatrix}.
\end{eqnarray}

(4) The 240$^o$-degree rotation $\widehat{C}_3^2$:
\begin{equation}\label{RT2}
\widehat{C}_3^2:\quad C_{\mathbf{i}\mu\sigma}\rightarrow C_{\mathbf{i}\nu\sigma}D^{(2)}_{\mu\nu}e^{-2i\sigma\pi/3},
\end{equation}
with
\begin{eqnarray}\label{asd2}
D^{(2)}_{\mu\nu}=\begin{pmatrix}
1&0&0\\
0&\frac{-1}{2} & \frac{-\sqrt{3}}{2}\\
0&\frac{\sqrt{3}}{2}& \frac{-1}{2}
\end{pmatrix}.
\end{eqnarray}
These symmetries bring constraint on the formula of SOC in the system. In the following, we first evaluate possible SOC conserving the SUS in (A), and then evaluate possible spin-flipping SOC in (B).

\subsection{Formula of spin-U(1)-symmetric SOC}
In this subsection, we analyze possible SOC terms with SUS, including the on-site formulae in 1 and the NN ones in 2.
\subsubsection{Formula of on-site SOC conserving SUS}
The on-site SOC with SUS can generally be written as $C^{\dagger}_{\mathbf{i}\mu\uparrow}C_{\mathbf{i}\nu\uparrow}\tilde{g}_{\mu\nu}+C^{\dagger}_{\mathbf{i}\mu\downarrow}C_{\mathbf{i}\nu\downarrow}\tilde{g}^{*}_{\mu\nu}+h.c.=
C^{\dagger}_{\mathbf{i}\mu\uparrow}C_{\mathbf{i}\nu\uparrow}(\tilde{g}_{\mu\nu}+\tilde{g}^{\dagger}_{\mu\nu})+C^{\dagger}_{\mathbf{i}\mu\downarrow}C_{\mathbf{i}\nu\downarrow}(\tilde{g}^{*}_{\mu\nu}+\tilde{g}^{T}_{\mu\nu})
\equiv C^{\dagger}_{\mathbf{i}\mu\uparrow}C_{\mathbf{i}\nu\uparrow}g_{\mu\nu}+C^{\dagger}_{\mathbf{i}\mu\downarrow}C_{\mathbf{i}\nu\downarrow}g^{*}_{\mu\nu}$, with the $g$-matrix to be Hermitian. Note that the TRS has been considered here. Let's evaluate the requirement on $g_{\mu\nu}$ by the $\widehat{C}_3^1$-rotation symmetry for the spin-up electrons.

As $C^{\dagger}_{\mathbf{i}\mu\uparrow}C_{\mathbf{i}\nu\uparrow}g_{\mu\nu}\xrightarrow{\quad \widehat{C}_3^1 \quad}C^{\dagger}_{\mathbf{i}\mu'\uparrow}C_{\mathbf{i}\nu'\uparrow}D^{(1)}_{\mu\mu'}D^{(1)}_{\nu\nu'}g_{\mu\nu}
=C^{\dagger}_{\mathbf{i}\mu\uparrow}C_{\mathbf{i}\nu\uparrow}(D^{(1)\dagger}gD^{(1)})_{\mu\nu}$. To keep rotation symmetry, we have
\begin{equation}\label{c31_sus}
D^{(1)\dagger}gD^{(1)}=g \to [D^{(1)\dagger},g]=0.
\end{equation}
Solving this equation, we get the symmetry-allowed formula for the $3\times3$ Hermitian matrix $g$ as
\begin{eqnarray}\label{g_mat}
g=\left(\begin{array}{ccc}
g_{1} & 0 & 0\\
0 & g_{2} & -i\gamma\\
0 & i\gamma & g_{2}
\end{array}\right).
\end{eqnarray}
Note that the weak diagonal term can be incorporated into the on-site chemical potential term, which will be ignored here. As a result, we get the Hamiltonian term describing this SOC,
\begin{eqnarray}
H^{(1)}_{SOC}=i\lambda_{so}\sum_{\mathbf{i}}\sigma(C^{\dagger}_{\mathbf{i}2\sigma}C_{\mathbf{i}3\sigma}-C^{\dagger}_{\mathbf{i}3\sigma}C_{\mathbf{i}2\sigma}).
\end{eqnarray}
It can be checked that this Hamiltonian satisfies all the symmetries listed above.

\subsubsection{Formula of NN- SOC with SUS}
The combined TRS represented by Eq.(\ref{TRS}), the mirror symmetry  represented by Eq.(\ref{MR}) and the Hermitian character of the Hamiltonian require that the NN- SOC conserving the SUS takes the formula of $C^{\dagger}_{\mathbf{i}\mu \uparrow}C_{\mathbf{i}+\mathbf{z}\nu \uparrow}g_{\mu\nu}+C^{\dagger}_{\mathbf{i}\mu \uparrow}C_{\mathbf{i}-\mathbf{z}\nu \uparrow}g_{\mu\nu}+C^{\dagger}_{\mathbf{i}\mu \downarrow}C_{\mathbf{i}+\mathbf{z}\nu \downarrow}g^{*}_{\mu\nu}+C^{\dagger}_{\mathbf{i}\mu \downarrow}C_{\mathbf{i}-\mathbf{z}\nu \downarrow}g^{*}_{\mu\nu}$, with the $g$-matrix Hermitian.

Then the $\widehat{C}_3^1$-rotation symmetry represented by Eq.(\ref{RT1}) requires the same formula Eq.(\ref{c31_sus}), which is solved as Eq.(\ref{g_mat}). Again, the weak diagonal part has extra spin-SU(2)-symmetry and can be incorporated into the band structure part, which will be ignored here. Therefore, we obtain
\begin{align}\label{SS1}
H^{(2)}_{SOC}=i\lambda_{so2}\sum_{\mathbf{i}}\sigma(&C^{\dagger}_{\mathbf{i}2 \sigma}C_{\mathbf{i}+\mathbf{z}3 \sigma}+C^{\dagger}_{\mathbf{i}2 \sigma}C_{\mathbf{i}-\mathbf{z}3 \sigma}\nonumber\\
&-C^{\dagger}_{\mathbf{i}3 \sigma}C_{\mathbf{i}+\mathbf{z}2 \sigma}-C^{\dagger}_{\mathbf{i}3 \sigma}C_{\mathbf{i}-\mathbf{z}2 \sigma})
\end{align}

\subsection{Formula of spin-flipping SOC}
In this subsection, we analyze possible spin-flipping SOC terms breaking SUS, including the on-site formulae in 1 and the NN ones in 2.
\subsubsection{Formula of spin-flipping on-site SOC }
It's proved here that the mirror-reflection symmetry $M$ will forbid spin-flipping on-site SOC.

Actually, assume we have a spin-flipping SOC term in the form of $C^{\dagger}_{\mathbf{i}\mu\sigma}C_{\mathbf{i}\nu\overline{\sigma }}g_{\mu\nu}$ with $\sigma=\uparrow$ or $\downarrow$, then from the mirror symmetry M about the plane passing through $\mathbf{i}$, we have: $C^{\dagger}_{\mathbf{i}\mu\sigma}C_{\mathbf{i}\nu\overline{\sigma }}g_{\mu\nu}\xrightarrow{\quad M \quad} C^{\dagger}_{\mathbf{i}\mu\sigma}C_{\mathbf{i}\nu\overline{\sigma }}\sigma\overline{\sigma }g_{\mu\nu}=-C^{\dagger}_{\mathbf{i}\mu\sigma}C_{\mathbf{i}\nu\overline{\sigma }}g_{\mu\nu}$. Since $M$ should be respected, we have $g_{\mu\nu}=0$, which suggests that the mirror-reflection symmetry $M$ will forbid spin-flipping on-site SOC.

\subsubsection{Formula of spin-flipping NN- SOC}
Here we evaluate the possible NN- spin-flipping SOC term. From the combination of the TRS represented by Eq.(\ref{TRS}), the mirror symmetry  represented by Eq.(\ref{MR}) and the Hermitian character of the Hamiltonian, such SOC term takes the following general formula,
\begin{align}\label{NN_spin_flipping}
H_{s-flip-SOC}=&\sum_{\mathbf{i}\mu\nu}
C^{\dagger}_{\mathbf{i}\mu\uparrow}C_{\mathbf{i}+\mathbf{z}\nu\downarrow}g_{\mu\nu}-
C^{\dagger}_{\mathbf{i}\mu\downarrow}C_{\mathbf{i}+\mathbf{z}\nu\uparrow}g^{*}_{\mu\nu}\nonumber\\
&-C^{\dagger}_{\mathbf{i}+\mathbf{z}\mu\uparrow}C_{\mathbf{i}\nu\downarrow}g_{\mu\nu}+
C^{\dagger}_{\mathbf{i}+\mathbf{z}\mu\downarrow}C_{\mathbf{i}\nu\uparrow}g^{*}_{\mu\nu},
\end{align}
with $g_{\mu\nu}=g_{\nu\mu}$. This Hamiltonian is already Hermitian.

Then the $\widehat{C}_3^1$-rotation symmetry represented by Eq.(\ref{RT1}) leads to the equation,
\begin{equation}\label{RT_S_flip}
D^{(1)\dagger}gD^{(1)}=ge^{\frac{2i\pi}{3}}.
\end{equation}
This equation can be solved as
\begin{equation}\label{g_flip}
g=\left(\begin{array}{ccc}
0 & i\lambda_{1} & \lambda_{1}\\
i\lambda_{1} & \lambda_{2} & i\lambda_{2}\\
\lambda_{1} & i\lambda_{2} & -\lambda_{2}
\end{array}\right),
\end{equation}
where $\lambda_1$ and $\lambda_2$ are two independent coupling constants.

Therefore, the symmetry-allowed NN- spin-flipping SOC term in the A$_2$Cr$_3$As$_3$ family takes the form of Eq.(\ref{NN_spin_flipping}), with the symmetric $g$-matrix provided by Eq.(\ref{g_flip}).

\section{Surface spectrum}\label{appendixC}

To calculate the surface spectrum, we first need a real-space BCS-MF Hamiltonian. For the pairing gap function introduced in the main text, only the $\mathbf{k}$-space gap function on the FS is provided, and thus we need a real-space pairing potential. Actually, the pairing state in K$_{2}$Cr$_{3}$As$_{3}$ can be approximately generated by the following MF Hamiltonian including only nearest-neighbor (NN) intra-orbital pairing potential\cite{Wu},
	\begin{eqnarray}\label{BCS}
	H&=&H_{\text{TB}}+H_{\text{SOC}}+H_{\Delta},\nonumber\\
	H_{\Delta}&=&\sum_{\mathbf{i}\mu}\left[\left(c^{\dagger}_{\mathbf{i}\mu\uparrow}c^{\dagger}_{\mathbf{i+z}\mu\downarrow}+
	c^{\dagger}_{\mathbf{i}\mu\downarrow}c^{\dagger}_{\mathbf{i+z}\mu\uparrow}\right)\Delta_{\mu}+h.c.\right].
	\end{eqnarray}
	This pairing state has a weak inter-band pairing component and its intra-band pairing component is well consistent with that obtained by the RPA approach. The topological invariants for this pairing state yield exactly the same results as those of the RPA.

\begin{figure}
	\includegraphics[width=1.0\columnwidth]{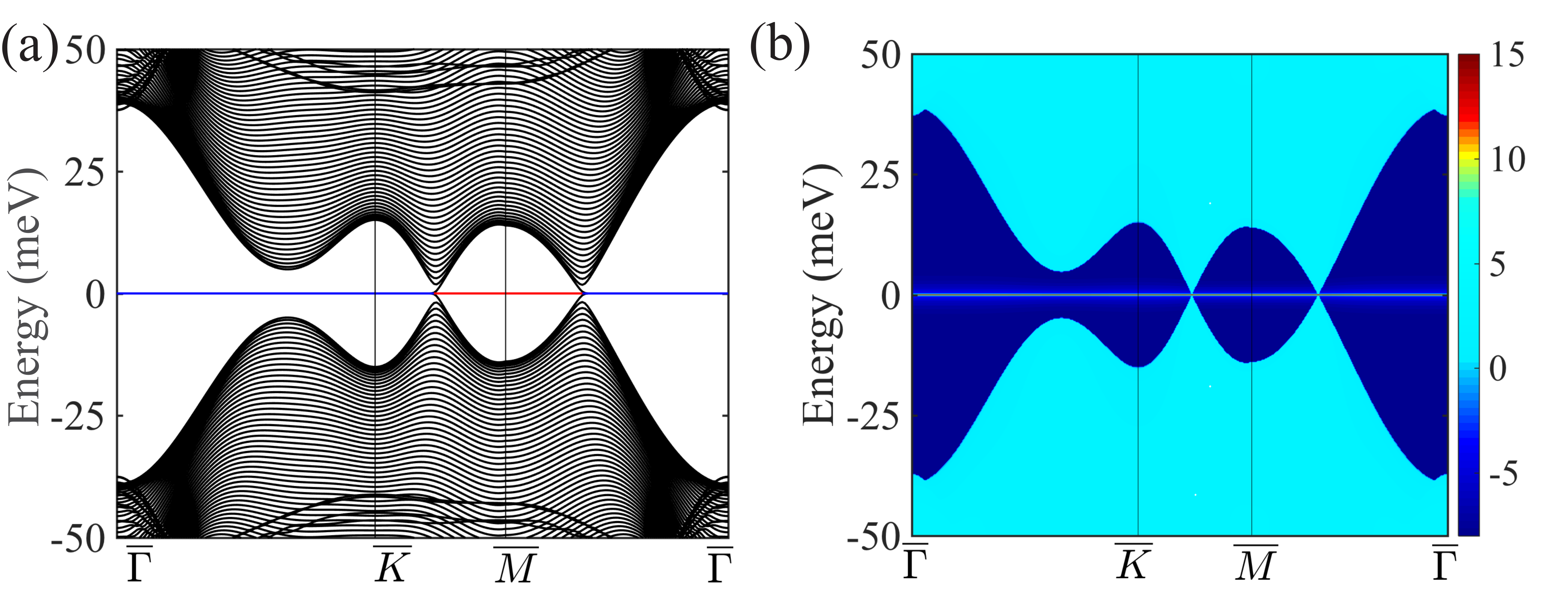}
	\caption
	{(Color online) The surface spectra obtained from (a) diagonalizing the BdG Hamiltonian directly using a slab geometry with open boundary condition along the $z$-axis (the width is 200c) and periodic ones along the $x$- or $y$- axis  and (b) the iterative Green's function approach for K$_{2}$Cr$_{3}$As$_{3}$ with SUS, respectively. The adopted $\Delta_{1}$=20 meV, $\Delta_{2}=\Delta_{3}$=40 meV are enhanced by an order of magnitude over realistic ones to enhance the visibility. The on-site SUS SOC $\lambda_{\text{so}}= 10$ meV.}
	\label{Surface_Nambu2}
\end{figure}

\begin{figure}
	\includegraphics[width=1.0\columnwidth]{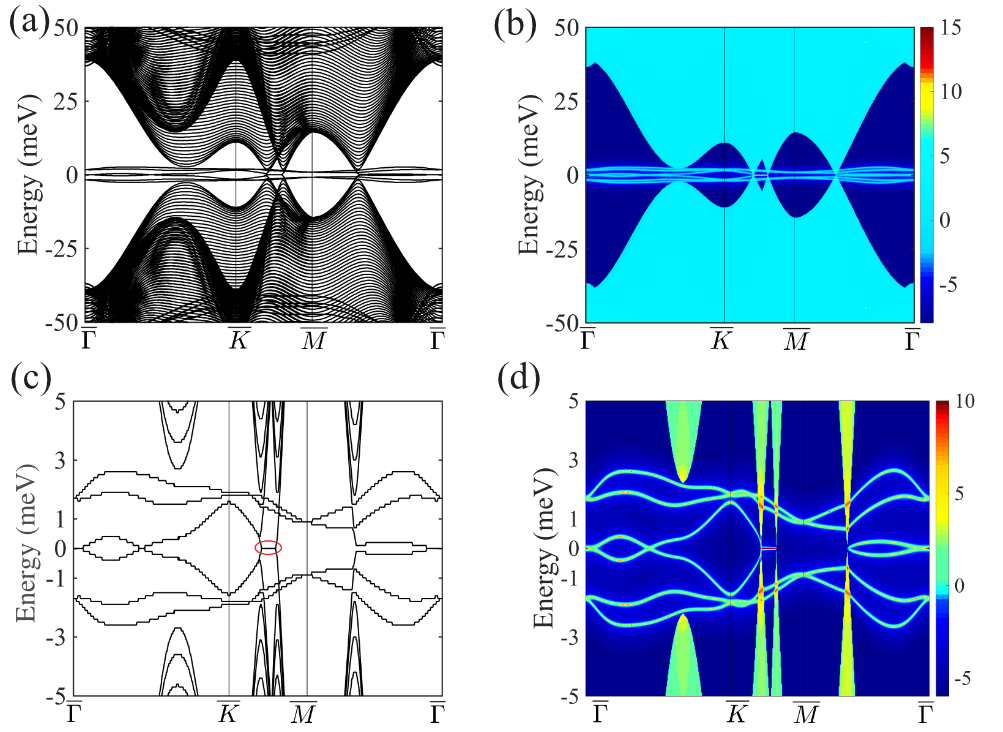}
	\caption
	{(Color online) The surface spectra obtained from (a) diagonalizing the BdG Hamiltonian directly using a slab geometry with open boundary condition along the $z$-axis (the width is 200c) and periodic ones along the $x$- or $y$- axis  and (b) the iterative Green's function approach for K$_{2}$Cr$_{3}$As$_{3}$ in the presence of spin-flipping SOC, respectively. (c) (d) are the zoom-in of (a) (b). The parameters are taken as $\Delta_{1}$=20 meV, $\Delta_{2}=\Delta_{3}$=40 meV, the on-site SUS SOC $\lambda_{\text{so}}= 10$ meV, the NN- SUS SOC $\lambda_{\text{so2}}= 2$ meV, and the NN-spin-flipping SOC $\lambda_{\text{1}}=\lambda_{\text{2}} =2$ meV.}
	\label{Surface_Nambu4}
\end{figure}

In the following, we adopt two different approaches to calculate the surface spectrum of the system described by Eq. (\ref{BCS}) in the presence of SUS or Rashba SOC. In the first approach, we directly diagonalize the Bogoliubov-de Genes Hamiltonian of the superconducting system using a slab geometry with open boundary condition along the $z$-axis (the width is 200c) and periodic ones along the $x$- or $y$- axis. In the second approach, we utilize an iterative method\cite{Sancho1985} to obtain the surface Green's functions of semi-infinite systems, from which we calculate the dispersions of the surface states. The results obtained from both approaches agree well with each other, which exhibit the exactly flat surface spectrums shown in Fig.~\ref{Surface_Nambu2} and Fig.~\ref{Surface_Nambu4}.

\section{STM}\label{appendixD}

The site-dependent differential conductance $dI/dV$ spectrum of the STM can be evaluated as
\begin{align}\label{STM1}
&\rho \left ( i_z,\omega  \right )=\rho \left ( i_x,i_y,i_z,\omega  \right )\\
&=-\text{Im}\sum_{E}\frac{\sum_{\mu \sigma } \left |  \left \langle E \left |  C_{i_x i_y i_z,\mu \sigma }^{\dagger}     \right | G \right \rangle\right |^{2}}{\omega -E+i0^{+}}\nonumber\\
&=-\text{Im}\sum_{E}\frac{\sum_{k_x, k_y, \mu \sigma } \left |  \left \langle E \left |  C_{k_x k_y i_z,\mu \sigma }^{\dagger}     \right | G \right \rangle  \frac{e^{-i(k_x i_x +k_y i_y)}}{\sqrt{N_x N_y}}  \right |^{2}}{\omega -E+i0^{+}}\nonumber\\
&=- \frac{2\text{Im}}{N_x N_y}\sum_{E}\frac{\sum_{k_x ,k_y, \mu, m } \left |  \left \langle E \left |  \gamma_{k_x,k_y,m }^{\dagger}     \right | G \right \rangle  \psi_{i_z \mu ,m}^{*}(k_x,k_y) \right |^{2}}{\omega -E+i0^{+}}\nonumber\\
&=- \frac{2\text{Im}}{N_x N_y}\sum_{k_x,k_y,m,\mu }\frac{ \left |  \psi_{i_z \mu ,m}(k_x,k_y) \right |^{2}}{\omega -E_{m}(k_x,k_y)+i0^{+}},
\end{align}
where $N_x, N_y$ are the size of lattice along x and y direction, $E_{m}(k_x,k_y)$ is the $m$-th eigenvalue of the BdG Hamiltonian matrix for the fixed momentum $(k_x,k_y)$ and $\psi_{i_z \mu ,m}(k_x,k_y)$ is the corresponding eigenvector. Note that the spectrum only depends on $i_z$ and not on $i_x$ or $i_y$. The coordinates $i_z=1(N_z)$ and $i_z=N_z/2$ correspond to the end and the middle of the sample, respectively.

\section{Fermi surface evolution and Lifshitz transition upon doping}\label{appendixE}

\begin{figure}
	\includegraphics[width=1.0\columnwidth]{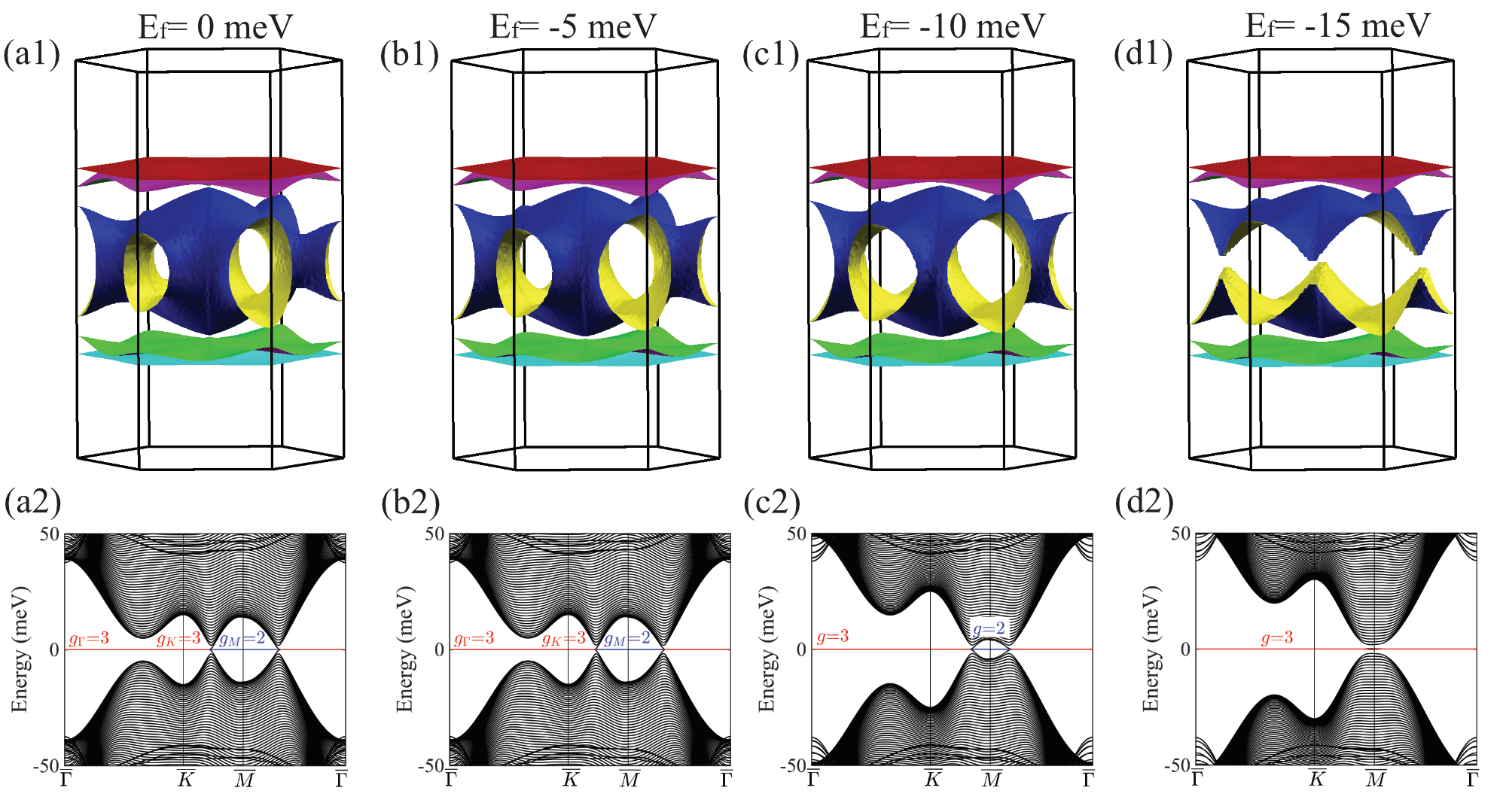}
	\caption
	{(Color online) (a1)-(d1) Fermi surfaces of the spin-up electrons for K$_{2}$Cr$_{3}$As$_{3}$ in the presence of the leading on-site SOC term for hole doping with different Fermi energy $E_f$= 0 meV, -5 meV, -10 meV, and -15 meV. (a2)-(d2) The corresponding surface spectrum. The segment marked red (blue) represents $g=3$ ($g=2$)  flat bands.}
	\label{FS_evolution1}
\end{figure}

\begin{figure}
	\includegraphics[width=1.0\columnwidth]{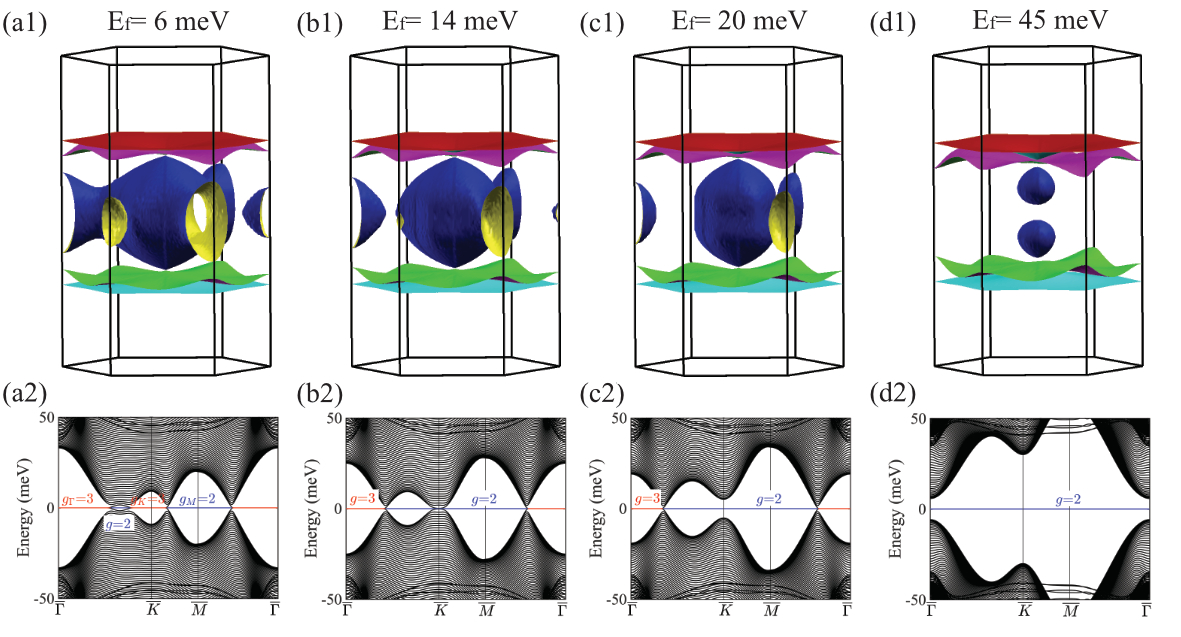}
	\caption
	{(Color online) Same as Fig.~\ref{FS_evolution1} but for electron doping with Fermi energy $E_f$= 6 meV, 14 meV, 20 meV, and 45 meV, respectively.}
	\label{FS_evolution2}
\end{figure}

The FS topology of  K$_2$Cr$_3$As$_3$ can be drastically changed upon slightly doping, accompanied by several Lifshitz transitions. As a result, the distribution of the number of topological flat surface bands will be easily engineered through doping, which can be detected by experiments. The FSs in both Figures ( Fig.~\ref{FS_evolution1} for hole doping and Fig.~\ref{FS_evolution2} for electron doping) are the FSs of the spin-up electrons, and the FSs for the spin down channel can be obtained by time-reversal operation. Consistent with the topological invariant calculations in the main text, the corresponding areas in the surface spectrum of Fig.~\ref{FS_evolution1} and Fig.~\ref{FS_evolution2} are covered by 2 or 3 flat bands on the surface Brillouin zone.

\end{document}